\documentclass[preprint,aps,showpacs,nofootinbib]{revtex4}

\usepackage{epsfig,amssymb,amsmath}

\def\bea{\begin{eqnarray}}
\def\eea{\end{eqnarray}}
\def\beq{\begin{equation}}
\def\eeq{\end{equation}}

\begin{document}
\draft
\tighten
\preprint{TU-907}
\title{\large \bf
       Singlet-doublet Higgs mixing and its implications on the Higgs mass
       in the PQ-NMSSM
       }
\author{
    Kwang Sik Jeong\footnote{email: ksjeong@tuhep.phys.tohoku.ac.jp},
    Yutaro Shoji\footnote{email: yshoji@tuhep.phys.tohoku.ac.jp},
    Masahiro Yamaguchi\footnote{email: yama@tuhep.phys.tohoku.ac.jp}}
\affiliation{
    Department of Physics, Tohoku University, Sendai 980-8578, Japan}
%\date{\today}

\vspace{2cm}

\begin{abstract}

We examine the implications of singlet-doublet Higgs mixing on the properties
of a Standard Model (SM)-like Higgs boson within the Peccei-Quinn invariant
extension of the NMSSM (PQ-NMSSM).
The SM singlet added to the Higgs sector connects the PQ and visible sectors
through a PQ-invariant non-renormalizable K\"ahler potential term, making the model
free from the tadpole and domain-wall problems.
For the case that the lightest Higgs boson is dominated by the singlet scalar,
the Higgs mixing increases the mass of a SM-like Higgs boson while reducing its
signal rate at collider experiments compared to the SM case.
The Higgs mixing is important also in the region of parameter space where the NMSSM
contribution to the Higgs mass is small, but its size is limited by the experimental
constraints on the singlet-like Higgs boson and on the lightest neutralino
constituted mainly by the singlino whose Majorana mass term is forbidden by
the PQ symmetry.
Nonetheless the Higgs mixing can increase the SM-like Higgs boson mass by
a few GeV or more even when the Higgs signal rate is close to the SM prediction,
and thus may be crucial for achieving
a 125 GeV Higgs mass, as hinted by the recent ATLAS and CMS data.
Such an effect can reduce the role of stop mixing.

\end{abstract}

%\pacs{}
\maketitle

%\tableofcontents
%\newpage

\section{Introduction}

A natural way to explain the smallness of the higgsino mass parameter in the minimal
supersymmetric standard model (MSSM) is to promote it to a SM singlet $S$ so that the
superpotential includes a coupling $\lambda S H_uH_d$.
This leads to the next-to-MSSM (NMSSM) \cite{Review-NMSSM}, which however generally
suffers from the tadpole \cite{Tadpole-problem1,Tadpole-problem2} and domain-wall
\cite{Domain-wall-problem1,Domain-wall-problem2} problems once one adds self-interactions
of $S$ to avoid a visible axion.
In particular, the tadpole problem makes it difficult to embed the NMSSM into
a grand unified theory.

We have recently pointed out in Ref. \cite{PQ-NMSSM} that all the problems arising
due to the SM singlet $S$ can be avoided when the NMSSM is extended to incorporate
the Peccei-Quinn (PQ) symmetry solving the strong CP problem
\cite{PQ-mechanism1,PQ-mechanism2}.
The PQ symmetry protects $S$ from acquiring large tadpoles.
Furthermore, the domain-wall problem is resolved by introducing an appropriate
number of PQ messengers or considering the situation where the saxion is displaced
far from the origin during inflation.
In the PQ-invariant extension of the NMSSM (PQ-NMSSM), $S$ plays the role of a
messenger that connects the PQ sector with the visible sector.
This is done through its non-renormalizable coupling with a SM singlet responsible
for spontaneous PQ symmetry breaking at a scale much higher than the electroweak scale.
Such non-renormalizable coupling generates a small effective tadpole for $S$.
As a result, the electroweak scale originates from the SUSY breaking scale and
the axion decay constant.
Another interesting property is the existence of a relatively light neutralino
dominated by the singlino.

The inclusion of $S$ modifies the Higgs and neutralino sectors, and opens the
possibility to have singlet-doublet Higgs mixing.
Such mixing increases the mass of the SM-like Higgs boson when the lightest
Higgs boson is singlet-like \cite{Mixing-NMSSM}.
It should be noted that the mixing effect is important even in the region with
large $\tan\beta$ and small $\lambda$, where the NMSSM tree-level contribution
to the Higgs quartic coupling and thus to the Higgs boson mass is negligible.
The PQ-NMSSM, in which quadratic and cubic terms in $S$ are forbidden by the PQ
symmetry, includes a light singlino-like neutralino.
This results in that the Higgs sector has a different phenomenology from other
NMSSM models because, if kinematically allowed, the Higgs bosons and the $Z$ boson
invisibly decay into a pair of neutralinos through couplings proportional
to $\lambda$.
Furthermore, the Higgs quartic coupling can receive additional sizable radiative
corrections involving the Yukawa interaction, Higgs-higgsino-singlino \cite{PQ-NMSSM}.
In this paper, we examine the implications of the Higgs mixing on the properties
of a SM-like Higgs boson by combining the experimental constraints placed on
the singlino-like neutralino and the lightest Higgs boson.

The recent ATLAS and CMS data hint the existence of a SM-like Higgs boson with
mass around 125 GeV \cite{Higgs-LHC}.
To account for this, one needs a particular mechanism of SUSY breaking giving large
stop mixing or an extension of the MSSM for superparticles having masses around
a TeV as suggested by the gauge hierarchy problem \cite{SUSY}.\footnote{
See Refs. \cite{Recent-NMSSM1,Kobayashi:2012xv,Recent-NMSSM2,Recent-NMSSM3,Mirage-NMSSM,NMSSM-models}
for a recent discussion of singlet extensions of the MSSM.
}
Thus, the singlet-doublet Higgs mixing may be crucial for achieving a 125 GeV Higgs
mass without invoking large stop mixing.
The Higgs mixing needs to be small in order for the Higgs signal rate at collider
experiments to be close to the SM prediction.
Nonetheless, we find that it can still increase the Higgs boson mass by more
than a few GeV while avoiding the experimental constraints.

In the next section, we briefly discuss the properties of the PQ-NMSSM
and the effects of the singlet-doublet Higgs mixing.
To see the impacts of the Higgs mixing, we construct a low energy effective theory
assuming that the MSSM superparticles except the higgsinos have masses around a TeV.
In section 3, we explore the properties of the singlino-like light neutralino,
on which the LEP experiments put constraints.
Finally, in section 4, we discuss how large the Higgs mixing can be in the PQ-NMSSM
and how much it can increase the mass of a SM-like Higgs boson.
Section 5 is devoted to discussion and conclusions.

\section{PQ-invariant extension of the NMSSM}

The PQ-NMSSM is obtained by extending the NMSSM to incorporate the PQ solution
to the strong CP problem, and the SM singlet $S$ added to the Higgs sector
connects the PQ and visible sectors through the operators,
\bea
\int d^4\theta\, \kappa \frac{X^{*2}}{M_{Pl}}\,S + \int d^2\theta
\lambda S H_uH_d,
\eea
where the PQ symmetry is assumed to be spontaneously broken by another SM
singlet $X$ at a scale much higher than the electroweak scale.
The $\kappa$ term induces an effective tadpole for $S$ at $\sim F^2_a/M_{Pl}$
after PQ symmetry breaking and drives $S$ to acquire a vacuum expectation value.
Here $F_a$ is the axion decay constant, and we drop a PQ-invariant superpotential
term $X^2H_uH_d/M_{Pl}$ since it can always be absorbed into the $\kappa$ term
by redefining $S$.
It should be noted that, protected by the PQ symmetry, $S$ does not have
a large tadpole.\footnote{
The tadpole problem can be avoided also by imposing a discrete $R$ symmetry
as discussed in Refs. \cite{nMSSM1,nMSSM2,nMSSM3,discrete-R}.
Then, the SM singlet $S$ added to the Higgs sector only has a superpotential
term $SH_uH_d$ and a small tadpole term induced after SUSY breaking.
See also Ref. \cite{PQ-NMSSM2}, where the NMSSM is extended to incorporate
the PQ symmetry and local $R$ symmetry.
}
On the other hand, as in other NMSSM models, the $\lambda$ term replaces
a supersymmetric $\mu$ term of the MSSM.
Having non-negligible coupling $\lambda$ to the Higgs doublets, $S$ modifies
the Higgs and neutralino sectors in a significant way.

\subsection{Low energy effective action}

The scalar potential for the extended Higgs sector depends on
the effective tadpole term of $S$ and the SUSY breaking parameters of
$S$ and $H_{u,d}$,
\bea
\label{Higgs-V}
V &=& \frac{g^2+g^{\prime 2}}{8} (|H_u|^2-|H_d|^2)^2
+ \frac{g^2}{2}|H^\dagger_u H_d|^2
%\nonumber \\
%&&
+ \left| \lambda H_uH_d + m^2_0 \right|^2
+ |\lambda S|^2 (|H_u|^2 + |H_d|^2 )
\nonumber \\
&&
+\, m^2_{H_u}|H_u|^2 + m^2_{H_d}|H_d|^2 + m^2_S |S|^2
+ \left(
A_\lambda \lambda S H_uH_d
- B_\kappa m^2_0 S
+ {\rm h.c.} \right),
\eea
where we take the same notation as in our previous paper \cite{PQ-NMSSM}, with
the exception that we will use $B$ and $\mu$ instead of $B_{\rm eff}$ and
$\mu_{\rm eff}$, respectively.
It should be noted that $m^2_i$, $A_\lambda$ and $B_\kappa$ are set
by the SUSY breaking scale $m_{\rm soft}$, while $m^2_0 \sim \kappa\, m_{\rm soft}
\langle|X|\rangle^2/M_{Pl}$
is determined by the axion decay constant.
Assuming for simplicity that the phases of $A_\lambda$ and $B_\kappa$ are aligned,
one can always make $A_\lambda$, $B_\kappa$, $\lambda$ and $m^2_0$ real
positive by rotating the phases of $S$ and $H_{u,d}$.
We take such a field basis throughout this paper.
Using the extremum conditions, one can replace $(B_\kappa,m^2_{H_u},m^2_{H_d},m^2_0)$
by $(B,\mu,\tan\beta,v)$, where $\tan\beta$ is the ratio between Higgs
vacuum expectation values with $v\simeq 174$ GeV, and
$B$ is the soft parameter associated with the effective $\mu$ term,
$\mu=\lambda\langle |S| \rangle$.
Then, the Higgs sector is parameterized by
\bea
(\lambda,m^2_S,A_\lambda,B,\mu,\tan\beta),
\eea
in terms of which $m^2_0$ is written $m^2_0=(B-A_\lambda)\mu/\lambda$,
implying that $B>A_\lambda$ in the model.
See also the appendix for the relations between other SUSY breaking terms and
the above parameters.

The region of parameter space that we will investigate is such that the singlet scalar
is relatively light, around the electroweak scale.
To see the impacts of the singlet-doublet Higgs mixing, we further assume
that heavy Higgs scalars and MSSM superparticles except the higgsinos
obtain TeV masses.
Integrating out the heavy fields, one obtains a low energy effective theory, which reduces
to the SM with additional fields: the singlet complex scalar $S$, singlino $\tilde S$
and higgsinos $\tilde H_{u,d}$.
The effective scalar interactions read
\bea
\label{Higgs-potential}
V_{\rm eff} &=& \frac{\lambda_H}{2} (|H|^2-v^2 )^2
+ \left(\frac{m^2_S}{\lambda^2}+v^2 \right)|\lambda S-\mu|^2 + (|H|^2-v^2 )|\lambda S-\mu|^2
\nonumber \\
&&
+\, \frac{1}{2}\,(2\mu-A_\lambda \sin2\beta) \Big\{ (|H|^2-v^2 )
(\lambda S-\mu) + {\rm h.c.}\Big\}
\nonumber \\
&&
-\, \frac{\sin2\beta}{2B\mu}\left|f^*_{\rm mix}\sin^2\beta-f_{\rm mix}\cos^2\beta
\right|^2|H|^2,
\eea
where $H$ denotes the light Higgs doublet scalar in the decoupling
limit.
The last term in the potential results from the tree-level exchange of
the heavy Higgs scalars having mass $\simeq (2B\mu/\sin2\beta)^{1/2}\gg m_W$:
\bea
f_{\rm mix} = A_\lambda (\lambda S -\mu) + \frac{\sin2\beta}{4}
(g^2+g^{\prime 2}-2\lambda^2)(|H|^2-v^2),
\eea
which vanishes at the vacuum.
The derivation of $f_{\rm mix}$ is presented in the appendix.
As in the conventional NMSSM, the Higgs quartic coupling is given by
$\lambda_H(m_{\rm soft})=\frac{g^2+g^{\prime 2}}{4}\cos^22\beta
+\frac{\lambda^2}{2}\sin^22\beta$ at the tree-level, and receives
threshold corrections coming from stop loops.
The Yukawa interactions relevant to our discussion are
\bea
-{\cal L}_{\rm eff} =
y_t \bar t_R Q H^c
+ y_b \bar b_R Q H + y^\prime_s S \tilde H_u \tilde H_d
+ y^\prime_u H \tilde H_u \tilde S
+ y^\prime_d H^c \tilde H_d \tilde S + {\rm h.c.},
\eea
where $H^c=i\sigma_2 H^*$, and the couplings are given by
$y^\prime_s=\lambda$, $y^\prime_u=\lambda \cos\beta$ and $y^\prime_d=\lambda \sin\beta$
at the scale $m_{\rm soft}$.

\subsection{Higgs mixing}

In this subsection, we briefly review the effects of Higgs mixing in an extension
of the SM where the SM Higgs boson mixes with a SM singlet scalar.
The low energy effective theory (\ref{Higgs-potential}) belongs to this class of
models.
The CP-even neutral scalars $h$ and $s$, coming from $H$ and $S$ respectively, compose
the mass eigenstates $H_{1,2}$:
\bea
H_1 &=& s\cos\theta -h\sin\theta,
\nonumber \\
H_2 &=& s\sin\theta + h\cos\theta,
\eea
which have mass $m^2_{H_1}=M^2_{ss} - M^2_{hs}\tan\theta$ and
$m^2_{H_2} = M^2_{hh} + M^2_{hs}\tan\theta$ for the Higgs mixing angle fixed by
\bea
\tan2\theta = \frac{2M^2_{hs}}{M^2_{hh}-M^2_{ss}},
\eea
where $M^2_{ij}=\langle \partial_i\partial_j V \rangle$ is the mass squared matrix
element for $(h,s)$, and should satisfy the stability conditions $M^2_{hh,ss}>0$ and
$(M^2_{hs})^2<M^2_{hh}M^2_{ss}$.

Let us consider the case where the lightest Higgs boson $H_1$ originates mainly
from the singlet $s$.
Then, the mass of a SM-like Higgs $H_2$
\bea
m^2_{H_2} = M^2_{hh} + (M^2_{hh}-m^2_{H_1})\tan^2\theta
\eea
receives a positive contribution from the mixing.
To see how much it can increase $m_{H_2}$, one should take into account the experimental
results on the Higgs search.
For a singlet-like $H_1$ lighter than 114 GeV, LEP places stringent constraints
on $e^+e^-\to ZH_1\to Zb\bar b$, i.e. on the effective coupling \cite{LEP-Higgs},
\bea
R^{b\bar b}_{H_1} \equiv \frac{g^2_{ZZH_1}}{g^2_{ZZh}}\, {\rm Br}(H_1\to b\bar b)
=  {\rm Br}(H_1\to b\bar b)\sin^2\theta,
\eea
where ${\rm Br}(H_1\to b\bar b)$ is the branching ratio of the corresponding process.
On the other hand, the production of $H_2$ at hadron
colliders proceeds essentially through the same processes as those in the SM.
However, the Higgs mixing modifies the discovery reach:
\bea
\label{H2-production-rate}
R^{\rm SM}_{H_2} \equiv \frac{\sigma(H_2)}{\sigma_{\rm SM}(h)}\,
{\rm Br}(H_2 \to {\rm SM})
= {\rm Br}(H_2 \to {\rm SM}) \cos^2\theta,
\eea
which provides the signal rate for the decays of $H_2$ to SM particles
in comparison with the SM case.
Here $\sigma(H_2)$ is the cross section for the $H_2$ production, while
$\sigma_{\rm SM}(h)$ is the Higgs production cross section in the SM.
If the Higgs signal rate is measured to be close to the SM prediction,
only small mixing is allowed.

The contribution to $m_{H_2}$ from the singlet-doublet Higgs mixing is
estimated by
\bea
\Delta m_{H_2} \equiv m_{H_2} - M_{hh}
= \frac{M^2_{hh}-m^2_{H_1}}{2 M_{hh}}\tan^2\theta
+ {\cal O}\left(\frac{\Delta m^2_{H_2}}{M_{hh}}\right),
\eea
in which $\tan\theta$ should be small enough to satisfy the LEP limits
on $R^{b\bar b}_{H_1}$, and will be further constrained if one requires
the $H_2$ signal rate to be close to what the SM predicts.
For $H_2$ having mass around 125 GeV, if for instance $R^{\rm SM}_{H_2}> 0.7$
is imposed, the contribution to $m_{H_2}$ from the Higgs mixing can be as large
as about 7 GeV when $H_1$ has mass in the range between about 90 and 100 GeV.
Here we have used that the LEP limits on $R^{b\bar b}_{H_1}$ require $\sin^2\theta$
less than about 0.25 for $H_1$ having mass in the indicated range, and that
$R^{\rm SM}_{H_2}>0.7$ translates into $\sin^2\theta<0.3$ assuming that $H_2$ dominantly
decays into SM particles.
Even if $R^{\rm SM}_{H_2}$ is to be more close to unity, for instance larger than
0.9, $\Delta m_{H_2}$ can be about 3 GeV at $m_{H_1}$ around 90 GeV.
The contribution from the Higgs mixing can thus be crucial for achieving a 125 GeV Higgs
mass in the supersymmetric SM.
In the PQ-NMSSM, the situation changes a bit due to invisible Higgs decays into
neutralinos, but not significantly if one requires $R^{\rm SM}_{H_2}\gtrsim 0.7$.
We will return to this point in section \ref{Higgs-sector}.

\section{Singlino-like neutralino LSP}

Having no Majorana mass term, the singlino obtains a small mass through
mixing with the neutral higgsinos, and becomes the dominant component
of the lightest neutralino $\chi^0_1$.
This opens the possibility that invisible decays of the $Z$ and Higgs bosons
into neutralinos are kinematically accessible.
Here we assume $R$-parity conservation.
Then, the size of the higgsino component in $\chi^0_1$ is severely constrained
by the LEP limit on the invisible $Z$ decay rate.
Also, the Higgs production rate at collider experiments is modified.

Let us examine the properties of the lightest neutralino before moving
to discuss the Higgs properties.
We note that the neutralino properties discussed here are independent
of the details of the Higgs sector.
In the low energy effective theory, the neutralino sector is comprised
of two neutral higgsinos and the singlino,
\bea
\tilde H^0_d = N_{3i} \chi^0_i, \quad
\tilde H^0_u = N_{4i} \chi^0_i, \quad
\tilde S = N_{5i} \chi^0_i,
\eea
where $i=(1,2,3)$, and we are assuming that gauginos are heavy and decouple
below the SUSY breaking scale.
The lightest neutralino couples to the SM particles through its higgsino
component, whose size depends on $\epsilon\equiv\lambda v/\mu$ and $\tan\beta$.
The relevant neutralino interactions are
\bea
-{\cal L} =
\frac{1}{2}y^s_{ij} \bar \psi^0_i \psi^0_j s
+ \frac{1}{2} y^h_{ij} \bar \psi^0_i \psi^0_j h
%+ y^\prime_h h \chi^0_1 \chi^0_2
+ \frac{1}{2} g_{ij} \bar{\psi}^0_i \gamma^\mu \gamma^5 \psi^0_j Z_\mu,
\eea
where $s$ and $h$ are the CP-even neutral Higgs boson coming from $S$ and $H$,
respectively, and $(\psi^0_i)^T=(\chi^0_i,\bar{\chi}^0_i)$ is the four-component
spinor.
The neutralino couplings read \cite{SUSY},
\bea
y^s_{ij} &=& \frac{\lambda}{\sqrt2} N_{3i}N_{4j}+ (i\leftrightarrow j),
\nonumber \\
y^h_{ij} &=& \frac{\lambda}{\sqrt2}
(N_{3i}N_{5j}\sin\beta +N_{4i}N_{5j}\cos\beta ) + (i\leftrightarrow j),
\nonumber \\
g_{ij} &=& \frac{g}{2\cos\theta_W}
( N_{3i}N^*_{3j}- N_{4i}N^*_{4j})  + (i\leftrightarrow j),
\eea
at the scale $m_{\rm soft}$.

For $\epsilon=\lambda v/\mu \ll 1$, we find the neutralinos to be
\bea
\left(%
\begin{array}{c}
  \chi^0_3 \\
  \chi^0_2 \\
  \chi^0_1 \\
\end{array}%
\right)
\simeq
\left(%
\begin{array}{ccc}
  \frac{1}{\sqrt2}-\frac{1+\sin2\beta+\cos2\beta}{4\sqrt2}\epsilon^2 &
  \frac{1}{\sqrt2}-\frac{1+\sin2\beta-\cos2\beta}{4\sqrt2}\epsilon^2 &
  \frac{\sin\beta+\cos\beta}{\sqrt2}\epsilon \\
  -\frac{1}{\sqrt2}+\frac{1-\sin2\beta+\cos2\beta}{4\sqrt2}\epsilon^2 &
  \frac{1}{\sqrt2}-\frac{1-\sin2\beta-\cos2\beta}{4\sqrt2}\epsilon^2 &
  \frac{\sin\beta-\cos\beta}{\sqrt2}\epsilon \\
  -\epsilon\cos\beta  & -\epsilon\sin\beta & 1-\frac{\epsilon^2}{2} \\
\end{array}%
\right)
\left(%
\begin{array}{c}
  \tilde H^0_d \\
  \tilde H^0_u \\
  \tilde S \\
\end{array}%
\right),
\eea
with masses given by
\bea
m_{\chi^0_1} &\simeq& \epsilon^2\mu (1-\epsilon^2)\sin2\beta,
\nonumber \\
m_{\chi^0_2} &\simeq& \mu\left( 1+\frac{1-\sin2\beta}{2}\epsilon^2 \right),
\nonumber \\
m_{\chi^0_3} &\simeq&  m_{\chi^0_1}+m_{\chi^0_2},
\eea
where we have ignored higher order terms in $\epsilon$ and small mixing
with the gauginos.
Thus, $\chi^0_1$ is the lightest superparticle (LSP) for small $\epsilon$,
and becomes lighter as $\tan\beta$ grows.

For gauginos with mass much larger than $\mu$, the lightest chargino
$\chi^\pm_1$ originates mainly from the charged higgsino.
Thus, the LEP bound on the chargino mass demands $\mu \gtrsim 100\,{\rm GeV}$.
In the case under consideration, the next-to-LSP (NLSP) is either
$\chi^0_2$ or the higgsino-like chargino $\chi^\pm_1$.
The 1-loop correction to the chargino mass is dominated by that from
gauge boson loops, which is about a few hundred MeV \cite{Mizuta:1992ja}.
As being higgsino-like, the chargino dominantly decays through the interaction
$W^-\chi^+_1\chi^0_1$ for $\mu>m_W+m_{\chi^0_1}$.
On the other hand, the main decay modes of $\chi^0_2$ are into $Z\chi^0_1$
and into a Higgs boson plus $\chi^0_1$, depending on the size of $\mu$.
% decays of $\chi^0_2$ into SM particles and the LSP,
%$\chi^0_2\to X_{\rm SM} \chi^0_1$, take place mainly with $X_{\rm SM}$ being $Z$ or
%a Higgs boson depending on the size of $\mu$.
For $\epsilon\ll 1$, the singlino-like neutralino couples very weakly to
other superparticles, and therefore supersymmetric cascade decays will proceed
first into the NLSP through the same interactions as in the MSSM \cite{Das:2012rr}.
The NLSP then promptly decays into the LSP.

Now we look into the constraints placed on the neutralino sector.
If $\chi^0_1$ has a mass smaller than half of the $Z$-boson mass,
the coupling $g_{11}$ should be sufficiently small in order not to exceed
the experimental bound on the invisible decay rate of $Z$ \cite{LEP-Z-inv,PDG}.
This requires
\bea
\label{inv-Z-decay}
\Gamma_{Z\to \chi^0_1\chi^0_1} =
\frac{g^2_{11}}{96\pi}m_Z \left(1-\frac{4m^2_{\chi^0_1}}{m^2_Z} \right)^{3/2}
\lesssim 2\,{\rm MeV},
\eea
which translates into
\bea
\left(\frac{\epsilon}{0.3}\right)^4\cos^2 2\beta
\,\lesssim 1.5,
\eea
for $\epsilon\ll 1$.
The above constraint becomes strong at large $\tan\beta$,
where we thus need a small $\lambda$ or large $\mu$.
For $\mu\gtrsim 100$ GeV as required by the chargino mass bound, the decay
$Z\to\chi^0_2\chi^0_1$ is kinematically forbidden.

\begin{figure}[t]
\begin{center}
\begin{minipage}{15cm}
\centerline{
{\hspace*{0cm}\epsfig{figure=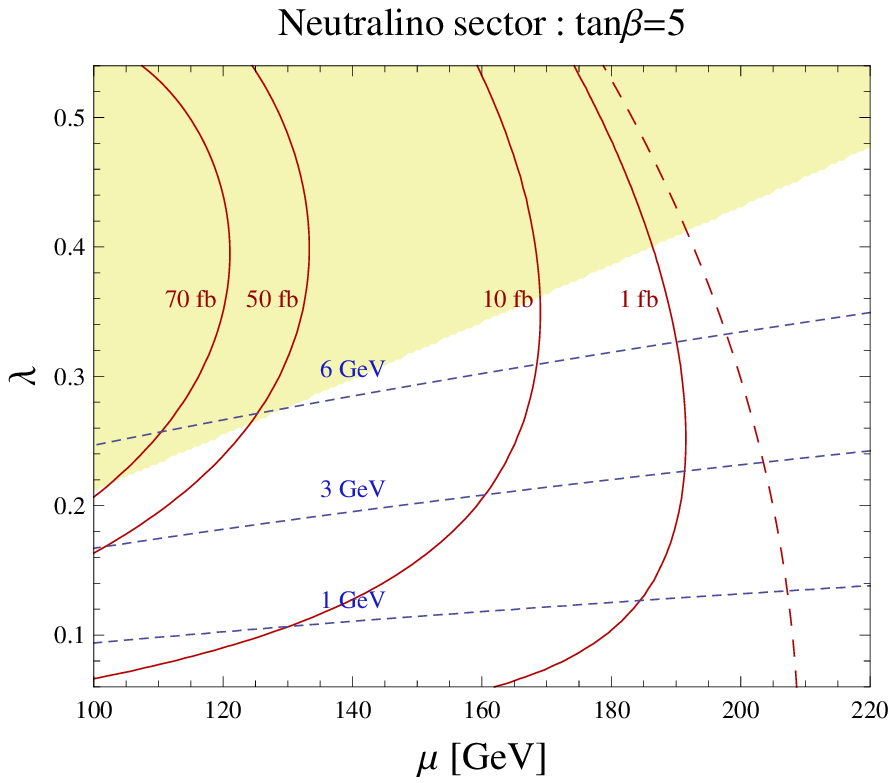,angle=0,width=7.4cm}}
{\hspace*{.2cm}\epsfig{figure=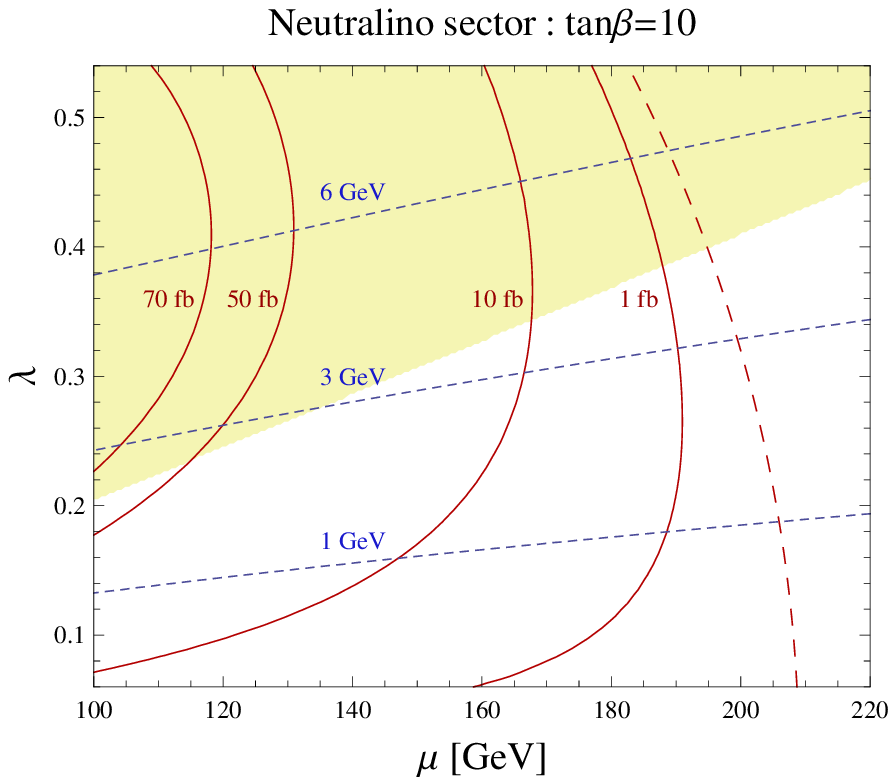,angle=0,width=7.4cm}}
}
\caption{Neutralino properties for $\tan\beta=5$ (left) and $\tan\beta=10$ (right).
The shaded region is excluded by the LEP bound on the invisible $Z$ decay rate.
Red lines represent the contours of the cross section for neutralino pair production
$\sigma(e^+e^-\to \chi^0_2\chi^0_1)$ at $\sqrt{s}=209$ GeV in the $(\mu,\lambda)$ plane.
In the right side of the dashed red line, $m_{\chi^0_2}+m_{\chi^0_1}$ is larger than
209 GeV.
We also show the contours for the mass of $\chi^0_1$ in the dashed blue line.
}
\label{fig:neutralino}
\end{minipage}
\end{center}
\end{figure}

LEP has placed constraints also on the neutralino production rate \cite{Dreiner:2009ic}.
Having higgsino components, neutralinos are produced at $e^+e^-$ colliders
through $Z$ exchange in the $s$ channel.
For $\mu$ much smaller than the neutral gaugino masses, production
processes through the exchange of selectron in the $t$ and $u$ channels
are negligible because of a small coupling for
the higgsino-electron-selection Yukawa interaction.
These processes are further suppressed when the sleptons are heavy.
The LEP bound on $e^+e^-\to \chi^0_2\chi^0_1$ requires
\bea
\label{LEP-neutralino}
\sigma(e^+e^-\to \chi^0_2\chi^0_1)\times {\rm Br}(\chi^0_2\to Z\chi^0_1)
\lesssim 70\,{\rm fb},
\eea
putting constraint on the size of $g_{11,12}$.
If $\mu$ is large enough to kinematically allow the decays of $\chi^0_2$ into
a Higgs boson and $\chi^0_1$, the branching fraction for
$\chi^0_2 \to Z\chi^0_1$ is reduced.

Fig. \ref{fig:neutralino} illustrates the properties of the neutralino
sector, which depend on $(\lambda,\mu,\tan\beta)$.
The shaded region, where $\Gamma_{Z\to\chi^0_1\chi^0_1}$ is larger than 2 MeV,
is excluded by the LEP bound on the invisible $Z$ decay rate.
In the allowed region with $\tan\beta\gtrsim 5$ and
$100\,{\rm GeV}\lesssim\mu\lesssim200\,{\rm GeV}$, we find the following properties.
The lightest neutralino has mass less than a few GeV, and the neutralino production
cross section $\sigma(e^+e^-\to \chi^0_2\chi^0_1)$ is less than 70 fb in
the non-shaded region.
The constraint (\ref{LEP-neutralino}) is thus evaded in that region regardless
of how strong the process $\chi^0_2$ into a Higgs boson and $\chi^0_1$ is.
On the other hand, if one takes $\tan\beta\lesssim 3$ and $\lambda\gtrsim 0.6$,
it is possible to render $\chi^0_1$ to have a mass larger than $m_Z/2$.

We close this section by briefly discussing the LSP relic abundance in the case
with $\epsilon\ll 1$ and $\mu$ around a few hundred GeV.\footnote{
See  Refs. \cite{DM-nMSSM,PH-nMSSM} for the case with not so small $\epsilon$.
In such case, $\chi^0_1$ obtains a mass larger than a few ten GeV, and its thermal relic
abundance can be consistent with the observed amount of dark matter.}
Since $\chi^0_1$ is then much lighter than the $Z$ and Higgs bosons and
interacts very weakly with them, its annihilation processes mediated by the exchange
of $Z$ or Higgs bosons are too weak to avoid overclosing the universe.
In order to resolve this cosmological difficulty, one may consider a superparticle lighter
than $\chi^0_1$, such as the axino or the gravitino, into which $\chi^0_1$ decays.
For those two, the relevant interactions include
\bea
\label{LSP-int}
{\cal L} =
C_{\tilde a}\, \tilde a \sigma^\mu \bar\chi^0_1 (\partial_\mu a)
+ C_{3/2}\, \tilde G_{1/2} \sigma^\mu \bar \chi^0_1 (\partial_\mu a) + {\rm h.c.},
\eea
with couplings given by
\bea
C_{\tilde a} &\sim& \frac{\langle K_{\bar X S} \rangle}{F_a} \sim \frac{\kappa}{M_{Pl}},
\nonumber \\
C_{3/2} &\sim& \frac{m_{\chi^0_1}}{m_{3/2}}\frac{1}{M_{Pl}},
\eea
where $\tilde a$ is the axino, and $a$ is the axion.
The $C_{\tilde a}$ interaction comes from the $\kappa$ term, while the other is the effective
interaction for the goldstino component $\tilde G_{1/2}$ of the gravitino
\cite{effective-gravitino-interaction}
for the case that the gravitino mass $m_{3/2}$ is smaller than $m_{\chi^0_1}$.
If the axino or the gravitino is the LSP, $\chi^0_1$ will decay with a very long lifetime.
Such late decay of $\chi^0_1$ would cause another cosmological problem because the produced LSP
becomes a hot dark matter component, whose energy density is severely constrained by the CMBR
and structure formation \cite{CMB,hot-dark-matter}.
It would be possible to avoid the bound on the hot dark matter abundance for a tiny LSP mass.
Then, the LSP, produced not by the late decay of $\chi^0_1$, and the axion can
constitute the cold dark matter of the universe.

On the other hand, if $R$-parity is conserved and $\chi^0_1$ is the LSP,
to avoid the LSP overproduction, one needs a sufficiently low reheating temperature or
a mechanism of late entropy production.
Indeed, in the PQ-NMSSM, it is plausible that the saxion potential energy dominates
the universe as it has a very flat potential lifted by SUSY breaking.
Let us assume such a case and examine the LSP production processes.
In the absence of the $\kappa$ term, the main decay mode of the saxion $\sigma$ is generally
$\sigma\to a a$.
However, the $\kappa$ term induces the interactions of $\sigma$, as well as of $\tilde a$,
with the NMSSM particles through kinetic mixing of the scalars and fermions in $X$ and $S$:
\bea
\langle K_{\bar X S} \rangle \sim \kappa \frac{F_a}{M_{Pl}} \sim
\frac{m^2_0}{m_{\rm soft}F_a}.
\eea
The saxion couplings to SM particles are $\sim \lambda m_{\rm soft}/F_a$
for $m^2_0\sim m^2_{\rm soft}$, and are crucial for suppressing the branching ratio
of the saxion decay into axions \cite{thermal-inflation-axion}.
The LSP is produced by thermal and non-thermal processes.
For thermally produced LSPs not to overclose the universe, the saxion decay temperature
should be low enough.
On the other hand, non-thermal production proceeds through the interaction
$\sigma \chi^0_1\chi^0_1$ having coupling $\sim y^s_{11} \langle K_{\bar X S} \rangle$
where $y^s_{11} \sim \lambda \epsilon^2 \sin2\beta$.
Since the involved coupling is further suppressed by $\epsilon^2$, it would not be
difficult to make the produced LSP energy density small.
Meanwhile, if the saxion decay into the axino is kinematically allowed, one should
pay attention also to axinos directly produced by saxion decays.
The axino decays via the $C_{\tilde a}$ coupling in (\ref{LSP-int}), but mainly through
the Higgs-$\tilde a$-$\chi^0_{1,2}$ interactions if kinematically accessible.
If lighter than the Higgs bosons, the axino will dominantly decay to $\chi^0_1$ with
a very long lifetime.
Since the produced LSP then behaves like a hot dark matter, its energy density
should be small.
A more detailed analysis will be given elsewhere.

\section{\label{Higgs-sector}Higgs sector}

In this section, we explore quantitatively the implications of singlet-doublet
Higgs mixing on the mass and the production rate of a SM-like Higgs boson in the case
where the lightest Higgs boson is dominated by the singlet scalar.

\subsection{Higgs properties}

Let us first examine the Higgs properties within the effective low energy theory
below $m_{\rm soft}$.
From the Higgs potential (\ref{Higgs-potential}), it is straightforward to
get
\bea
\label{matrix-elements}
M^2_{hh} &=& 2\lambda_H v^2,
\nonumber \\
M^2_{ss} &=& m^2_S + \lambda^2 v^2 \left(1-\frac{A^2_\lambda}{2B\mu}
\sin2\beta \cos^22\beta \right),
\nonumber \\
M^2_{hs} &=& \lambda v (2\mu-A_\lambda \sin2\beta).
\eea
Here radiative contributions to $M^2_{hh}$ are easily evaluated by
solving the renormalization group (RG) running equation of the Higgs quartic
coupling $\lambda_H$ in the effective theory \cite{Higgs-mass-MSSM1,Higgs-mass-MSSM2}.
The RG running is affected by the singlet and neutralinos when they
are light.
On the other hand, the CP-odd neutral Higgs boson $A$ originates from the singlet
scalar and obtains mass,
\bea
\label{mass-A}
m^2_A = m^2_S + \lambda^2 v^2 \left(1-\frac{A^2_\lambda}{2B\mu}
\sin2\beta \right),
\eea
which differs from $M^2_{ss}$ due to the contribution from $f_{\rm mix}$.

The main purpose of this work is to explore if the model can accommodate
$H_2$ around 125 GeV for a singlet-like Higgs $H_1$ having mass above about 80 GeV,
below which, as we will see, only small mixing is allowed.
In such a case, the decays of $H_{1,2}$ into SM particles are dominated by those
into $b\bar b$ and $VV^*$ where $V=(W,Z)$.
Since the processes are mediated via the $h$ component, one finds the decay
rates to be
\bea
\Gamma_{H_2\to {\rm SM}} &=& \Gamma_{\rm SM}(m_{H_2})\cos^2\theta,
\nonumber \\
\Gamma_{H_1\to {\rm SM}} &=& \Gamma_{\rm SM}(m_{H_1})\sin^2\theta,
\eea
in which $\Gamma_{\rm SM}(m_{H_i})$ is the decay rate of $h\to{\rm SM}$ obtained for
a SM Higgs boson $h$ having mass $m_{H_i}$, in the limit of no mixing with $s$.
In addition, because the singlino-like neutralino is very light, invisible Higgs
decays are possible:
\bea
\Gamma_{H_2\to \chi^0_i\chi^0_j} &=&
\frac{k^{H_2}_{ij}}{8\pi}
(y^h_{ij}\cos\theta + y^s_{ij} \sin\theta)^2 m_{H_2},
\nonumber \\
\Gamma_{H_1\to \chi^0_i\chi^0_j} &=&
\frac{k^{H_1}_{ij}}{8\pi}
(y^h_{ij}\sin\theta - y^s_{ij} \cos\theta)^2 m_{H_1},
\eea
for $i\geq j$, where neglecting the mass of $\chi^0_1$ we have $k^{H_{1,2}}_{11}\simeq 1/2$
and $k^{H_{1,2}}_{21}\simeq (1-m^2_{\chi^0_2}/m^2_{H_{1,2}})^2$ if the corresponding process
is kinematically allowed.
The Yukawa couplings responsible for the Higgs decays into neutralinos read
\bea
y^h_{11} &=& \sqrt2\lambda (1+ {\cal O}(\epsilon))\epsilon \sin2\beta,
\quad
y^s_{11} = -\frac{\lambda}{\sqrt2}
(1+ {\cal O}(\epsilon))\epsilon^2 \sin2\beta,
\nonumber \\
y^h_{21} &=& \frac{\lambda}{2} (1+ {\cal O}(\epsilon))(\cos\beta-\sin\beta),
\quad
y^s_{21} = -\frac{\lambda}{2} (1+ {\cal O}(\epsilon))(\cos\beta-\sin\beta) \epsilon,
\eea
at the scale $m_{\rm soft}$.

The neutralino Yukawa couplings to $H_{1,2}$ show that the invisible Higgs decay
$H_{1,2}\to \chi^0_1\chi^0_1$ becomes weak at large $\tan\beta$.
We also note that $y^h_{21}$ is stronger than the bottom Yukawa
coupling as long as $\lambda$ is larger than about 0.1 and $\tan\beta$ is not
close to unity.
Hence, if kinematically open, the mode $H_i \to \chi^0_2\chi^0_1$,
followed by the decay of $\chi^0_2$ into $Z$ (or a Higgs boson)
and $\chi^0_1$, will dominate the Higgs decays while making Higgs searches
much more difficult.
Meanwhile, from the couplings $y^{h,s}_{11}$, it is straightforward to see
\bea
\label{Higgs-invisible-decay}
\frac{\Gamma_{H_i\to \chi^0_1\chi^0_1}}{\Gamma_{H_i\to b\bar b}}
=
\frac{c^2_{H_i}}{6} \left(
\frac{v}{m_b} \lambda \epsilon \sin2\beta
\right)^2,
\eea
where $c_{H_1}\simeq 2+\epsilon \cot\theta$ and
$c_{H_2}\simeq 2-\epsilon \tan\theta$ for $\epsilon\ll 1$, neglecting the masses
of the final states.
The branching ratio for the invisible Higgs decay is essentially determined by the
above quantity, which increases as $\tan\beta$ decreases and $\epsilon$ grows for
given $\tan\theta$.
The invisible channel of $H_1$ relaxes the LEP constraints on it, but that of
$H_2$ reduces the Higgs signal rate compared to the SM prediction.

\subsection{Effects of Higgs mixing}

Taking into account various constraints on the model parameters, we investigate
how much the Higgs mixing can contribute to the mass of the SM-like Higgs $H_2$,
which is assumed to be around 125 GeV as hinted by the recent ATLAS and CMS data
\cite{Higgs-LHC}.
The Higgs mixing should be small in order for the signal rate of $H_2$ decays
into SM particles to be near what is predicted by the SM.\footnote{
Large stop mixing can enhance the gluon-Higgs coupling mediated by squarks,
altering the rate of the gluon fusion for the Higgs production \cite{Harlander:2004tp}.
However, such effects are non-negligible only when stops are relatively light.
For sfermions having masses around a TeV or higher, one can take
$\sigma(H_2)/\sigma_{\rm SM}(h)\simeq \cos^2\theta$ in (\ref{H2-production-rate}).
}
Here we focus on the region with $R^{\rm SM}_{H_2}$ larger than 0.7.
In the analysis, it is convenient to use
\bea
(\lambda,\mu,\tan\beta,m_{H_1},M_{hh},\sin^2\theta)
\eea
instead of the parameters $(\lambda,\mu,\tan\beta,B,A_\lambda,m^2_S)$.
In the limit of $\lambda=0$, which corresponds to the MSSM case, large stop mixing
is required to realize $M_{hh}$ around 125 GeV for stops having a TeV mass.
The NMSSM contribution to the Higgs quartic coupling improves the situation, but only
in the low $\tan\beta$ regime if $\lambda$ is less than about 0.7 as required by the
perturbativity constraint.
It is thus interesting to consider the increase of $m_{H_2}$ by the Higgs mixing.

Before proceeding further, we summarize the region of parameter space under
investigation here:
\bea
\label{parameter-space}
&& 0.1 \lesssim \lambda \lesssim 0.4,
\quad
120\,{\rm GeV} \lesssim \mu \lesssim 300\, {\rm GeV},
\quad
5 \lesssim \tan\beta \lesssim 20,
\nonumber \\
&& 80\,{\rm GeV} \lesssim  m_{H_1} \lesssim 110\, {\rm GeV},
\quad
115\,{\rm GeV} \lesssim M_{hh} \lesssim 120\, {\rm GeV},
\quad
\sin^2\theta \lesssim 0.3,
\eea
where the MSSM superparticles except the higgsinos are assumed to have masses around
a TeV.
It has been taken into account that the constraint (\ref{inv-Z-decay})
requires a rather small $\lambda$ unless $Z\to \chi^0_1\chi^0_1$ is kinematically
forbidden, and that $\sin^2\theta$ less than 0.3 is needed to get $R^{\rm SM}_{H_2}>0.7$.
Furthermore, to ensure the stability of the electroweak vacuum $(M^2_{hs})^2<M^2_{hh}M^2_{ss}$,
one needs $A_\lambda \sim \mu\tan\beta$ unless $\lambda$ is very small.
This leads us to consider $A_\lambda$ around or above a TeV for $\mu$ at a few hundred GeV,
which is consistent with the assumption of the decoupling limit $(2B\mu/\sin2\beta)^{1/2}\gg m_W$
because $B>A_\lambda$ in the model.

Let us explain a bit more on the above parameter range.
We consider $\mu$ larger than about 120 GeV so that $H_2 \to \chi^0_2\chi^0_1$ is
kinematically closed, which would otherwise dominate the Higgs decay and reduce
the identification capability of $H_2$ signals at collider experiments.
For $\mu\gtrsim 120$ GeV and $\lambda\lesssim 0.4$, $\chi^0_1$ is lighter than about
10 GeV, and the invisible channel $H_2\to \chi^0_1\chi^0_1$ becomes strong at
low $\tan\beta$ while making Higgs searches difficult.
To avoid such a situation, we consider $\tan\beta$ larger than 5.
Finally, we take $\tan\theta>0$, which is the case with $2\mu>A_\lambda \sin2\beta$,
because in order to make the Higgs mixing effect sizable it is preferred that
the branching fraction of the invisible decay is large for $H_1$ but small for $H_2$
as much as possible.
This requires $\tan\theta>0$ as can be seen from (\ref{Higgs-invisible-decay}).

\begin{figure}[t]
\begin{center}
\begin{minipage}{15cm}
\centerline{
{\hspace*{0cm}\epsfig{figure=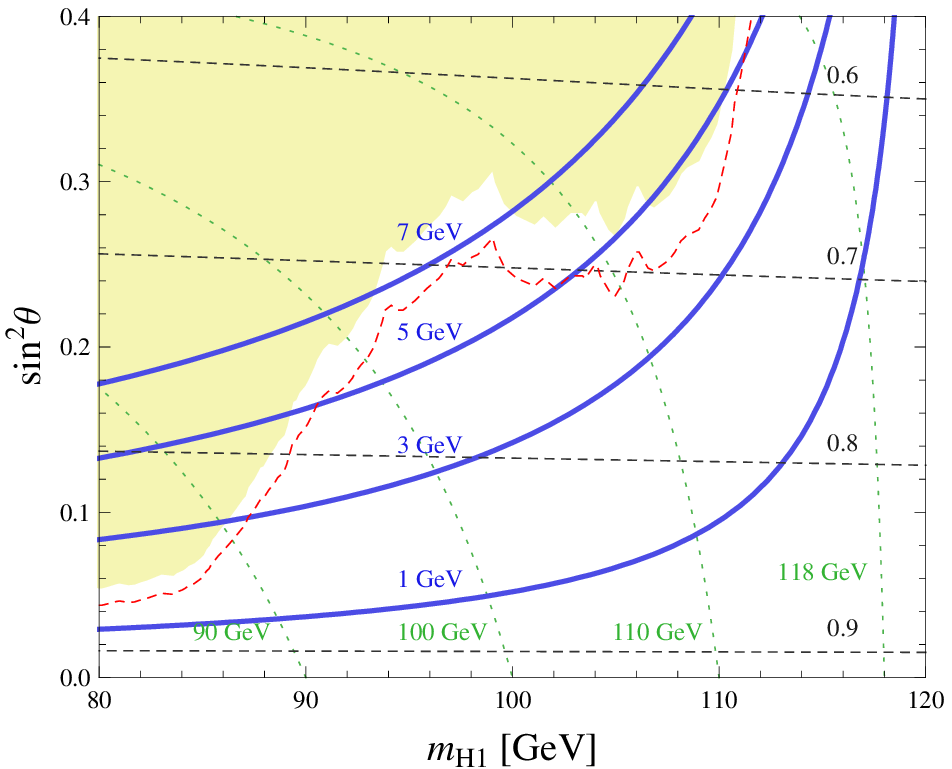,angle=0,width=7.4cm}}
{\hspace*{.2cm}\epsfig{figure=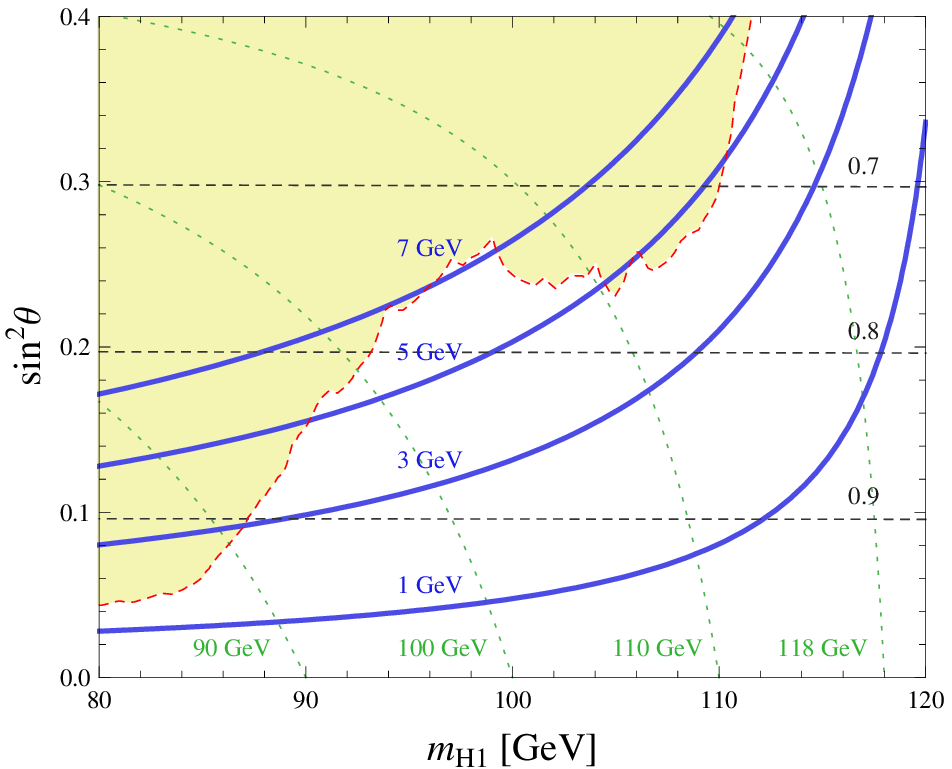,angle=0,width=7.4cm}}
}
\caption{Higgs mixing in two cases:
$(\lambda,\mu,\tan\beta,M_{hh})=(0.2,180\,{\rm GeV},10,120\,{\rm GeV})$
for the left panel, while $(\lambda,\mu,\tan\beta,M_{hh})=(0.1,130\,{\rm GeV},15,122\,{\rm GeV})$
for the right one.
We plot the constant contours of $\Delta m_{H_2}= m_{H_2}-M_{hh}$ in the blue line,
and also those of the mass of the CP-odd Higgs boson in the dotted greed line
on the $(m_{H_1},\sin^2\theta)$ plane.
The dashed black line represents the $H_2$ signal rate $R^{\rm SM}_{H_2}$, compared to
the SM case.
The shaded region is excluded by the LEP limits on the coupling of $H_1$ to $b\bar b$.
This constraint is a bit relaxed due to the invisible mode $H_1\to \chi^0_1\chi^0_1$.
For the comparison, we show the bound on $R^{b\bar b}_{H_1}$ in the dashed red line,
assuming that $H_1$ decays only into $b\bar b$.
}
\label{fig:Higgs}
\end{minipage}
\end{center}
\end{figure}

In Fig. \ref{fig:Higgs}, we illustrate the effects of the singlet-doublet Higgs mixing for
two cases.
For the case with $(\lambda,\mu,\tan\beta,M_{hh})=(0.2,180\,{\rm GeV},10,120\,{\rm GeV})$,
the neutralinos have masses $m_{\chi^0_2}\simeq 182.7$ GeV and $m_{\chi^0_1}\simeq 1.3$ GeV,
and the invisible $Z$ decay width into neutralinos is $\Gamma_{Z\to \chi^0_1\chi^0_1}\simeq 0.21$ MeV.
In this case, the Higgs mixing can increase the mass of $H_2$ upto about 7 GeV
in the range $R^{\rm SM}_{H_2}>0.7$ for $H_1$ having mass around 95 GeV.
On the other hand, for the other case with
$(\lambda,\mu,\tan\beta,M_{hh})=(0.1,130\,{\rm GeV},15,122\,{\rm GeV})$, we have
$m_{\chi^0_2}\simeq 131$ GeV and $m_{\chi^0_1}\simeq 0.3$ GeV.
The invisible $Z$ decay width into neutralinos is $\Gamma_{Z\to \chi^0_1\chi^0_1}\simeq 0.05$ MeV.
For this case, $\Delta m_{H_2}$ can be as large as 3 GeV at $m_{H_1}$ around 90 GeV even when
one requires $R^{\rm SM}_{H_2}$ larger than 0.9.
Therefore the Higgs mixing can considerably reduce the role of stop mixing in achieving
a 125 GeV Higgs boson mass.
Meanwhile, in both cases, the branching ratio for $H_2\to \chi^0_1\chi^0_1$ is smaller than 0.1,
and thus below the detectable level at the LHC \cite{Desai:2012qy,Cao:2012im}.

\begin{figure}[t]
\begin{center}
\begin{minipage}{15cm}
\centerline{
{\hspace*{0cm}\epsfig{figure=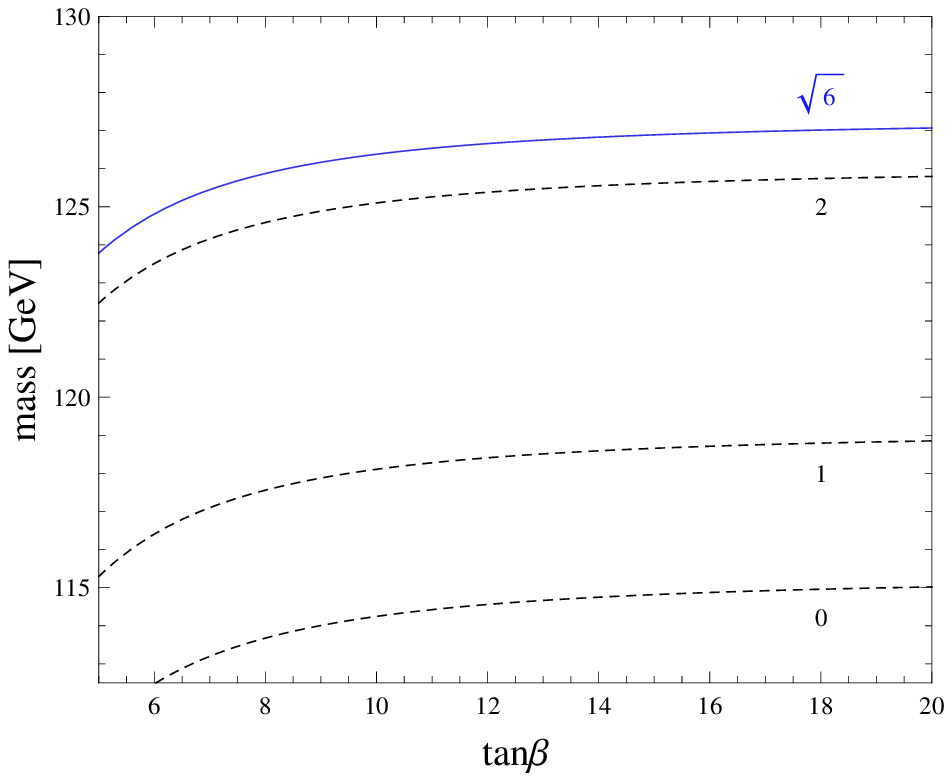,angle=0,width=7.4cm}}
{\hspace*{.2cm}\epsfig{figure=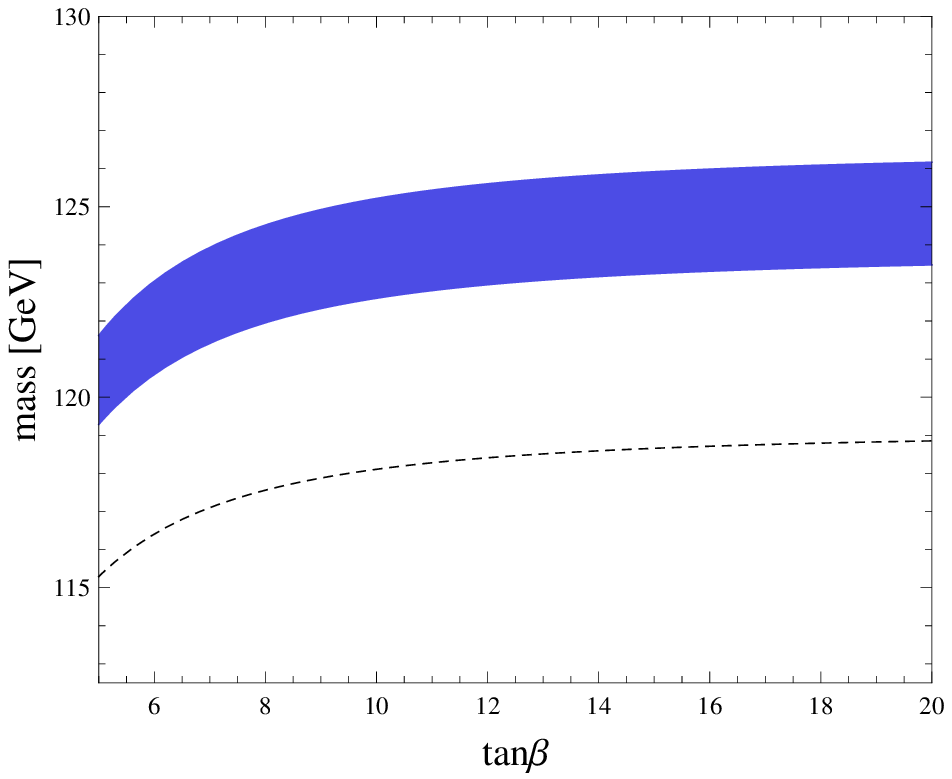,angle=0,width=7.4cm}}
}
\caption{Higgs boson mass for the case with $m_{\rm soft}=1.5$ TeV,
$\mu=180$ GeV and $\lambda=0.2$.
The left panel shows $M_{hh}$ for $X_t=0,1,2,\sqrt6$ from the below, respectively.
The case with $X_t=\sqrt6$ corresponds to the maximal stop mixing.
In the right panel, the blue band is $m_{H_2}$ obtained for $X_t=1$,
$m_{H_1}=95$ GeV and $0.18\leq\sin^2\theta\leq0.26$, for which
$R^{\rm SM}_{H_2}$ lies in the range between 0.73 and 0.8.
}
\label{fig:Mhh-mH2}
\end{minipage}
\end{center}
\end{figure}

In the low energy effective theory below $m_{\rm soft}$, $M_{hh}$ is
determined by the relation $M^2_{hh}=2\lambda_H v^2$ for $\lambda_H$ renormalized
at the electroweak scale.
The RG running of $\lambda_H$ is affected by the higgsinos for $\mu$ less than
$m_{\rm soft}$, but only slightly for small $\lambda$.
Note also that $M^2_{ss}$ and $M^2_{hs}$ explicitly depend on the SUSY breaking
parameters.
The left panel of Fig. \ref{fig:Mhh-mH2} shows the value of $M_{hh}$ for different
values of the stop mixing parameter $X_t\equiv (A_t-\mu\cot\beta)/m_{\rm soft}$
in the case with $m_{\rm soft}=1.5$ TeV, $\mu=180$ GeV and
$\lambda=0.2$.\footnote{
For simplicity, we use (\ref{matrix-elements}) to calculate $M^2_{hh}$ with
$\lambda_H$ evaluated by the use of one-loop RG equations, where we assume
the squarks/sleptons and gauginos to have a universal mass $m_{\rm soft}$.
This is sufficient for the purpose of our discussion on how much the
singlet-doublet Higgs mixing can increase the mass of the SM-like Higgs boson.
Note that higher-loop corrections to $M_{hh}$, which involve the strong
gauge coupling and the Yukawa couplings of the third generation fermions,
can induce a shift of a few GeV on the Higgs boson mass \cite{Two-loop-Higgs-mass}.
%There is also about 1 GeV uncertainty due to the top quark mass,
%$m_t=173.2\pm 1.8$ GeV for $m_{\rm soft}$ around TeV.
}
Here $A_t$ is the trilinear soft parameter for $H_u\tilde t_R \tilde Q_L$,
and the contribution from stop loops is maximized at $X_t=\sqrt 6$.
As one can see, $M_{hh}$ around 125 GeV requires the almost maximal stop mixing.
The situation does not change much at $\tan\beta\gtrsim 5$ if $\lambda$ is less
than about 0.7 as required for the interaction $SH_uH_d$ to remain perturbative
upto $M_{\rm GUT}\sim 10^{16}$ GeV.
In the right panel, the blue band represents the Higgs boson mass $m_{H_2}$ for
the case with $X_t=1$ and the same values of $m_{\rm soft}$, $\mu$ and $\lambda$.
Here we have taken $m_{H_1}=95$ GeV and $0.18\leq\sin^2\theta\leq0.26$,
which are obtained for $m_S\sim 90$ GeV and $A_\lambda \sim \mu \tan\beta \sim $
(a TeV).
In the indicated range of model parameters, $R^{\rm SM}_{H_2}$ has a value
between 0.73 and 0.8.
If one requires $R^{\rm SM}_{H_2}$ more close to unity, the amount of mass
increase by the mixing will be lowered.

On the other hand, the mass of the CP-odd Higgs boson $A$ reads
\bea
m^2_A =
M^2_{ss} - \lambda^2v^2 \frac{A^2_\lambda}{2B\mu}\sin^3 2\beta
= m^2_{H_1} + m^2_{H_2} - M^2_{hh}
- \lambda^2v^2 \frac{A^2_\lambda}{2B\mu}\sin^3 2\beta,
\eea
where we have used the relations (\ref{matrix-elements}) and (\ref{mass-A}).
In the parameter space under consideration, if the singlet-doublet mixing
yields $\Delta m_{H_2}$ of a few GeV or larger, $A$ has mass,
\bea
m_A \simeq m_{H_1} + \frac{m_{H_2}}{m_{H_1}}\,\Delta m_{H_2}
-|{\cal O}(1)|\times \frac{\lambda^2v^2}{m_{H_1}}\frac{1}{\tan^2\beta},
\eea
and thus is heavier than $H_1$ by about $\Delta m_{H_2}$, but always
lighter than $H_2$ for $m_{H_1}<M_{hh}$.
Fig. \ref{fig:Higgs} shows the dependence of $m_A$ on $m_{H_1}$ and
the singlet-doublet mixing.
Note that $H_2$ with mass around 125 GeV is forbidden to decay on-shell into
$H_1H_1$ or $AA$ for $80\,{\rm GeV}\lesssim m_{H_1} \lesssim 110\,{\rm GeV}$,
where $\Delta m_{H_2}$ can be sizable.
Meanwhile, in the decoupling regime, $A$ is mostly singlet-like and contains
only a small doublet component fixed by the mixing angle,
\bea
\theta^\prime \simeq \frac{\lambda v A_\lambda}{2|B\mu|}\sin2\beta,
\eea
to which all the couplings of $A$ to the SM fermions and gauge bosons are
proportional.
The $Ab\bar b$ coupling $\simeq (y_b\tan\beta)\theta^\prime$ is at most about
$\lambda y_b$ for $B>A_\lambda$, but can be similar
to or larger than the $A\chi^0_1\chi^0_1$ coupling $\simeq \lambda \epsilon^2\sin2\beta$
in the parameter space (\ref{parameter-space}).
There are no $AZZ$ and $AWW$ couplings at tree-level, while the $ZH_iA$ ($i=1,2$)
coupling is highly suppressed by small factors since it is generated from the
$Zh$-(doublet CP-odd Higgs boson) interaction whose coupling itself vanishes
in the decoupling limit.
One thus finds that $A$ mainly decays into $b\bar b$ and $\chi^0_1\chi^0_1$, but
evades the LEP constraints from $e^+e^-\to ZA \to Zb\bar b$ due to the suppressed
production cross section.
The process $e^+e^-\to Z^* \to H_1A$ (and $H_2A$ if kinematically allowed)
also produces $A$, but only with small rates due to the suppressed $ZH_1A$ coupling.
In addition, the decay of the SM-like Higgs boson via $H_2\to AA^* \to 4b$
is suppressed compared to $H_2\to ZZ^* \to 4b$ because the $Ab\bar b$ coupling is
smaller than $\lambda y_b$ and $H_2$ couples to $AA$ with coupling $\simeq \lambda^2 v$.

Finally, we discuss the effect of mixing between the doublet Higgs bosons, which
we have neglected since it is small when the heavier doublet Higgs obtains a mass
much larger than the electroweak scale.
In the MSSM, the light CP-even Higgs boson is composed out of $H^0_{u,d}$,
$h=(-{\rm Re}(H^0_d) \sin\alpha+{\rm Re}(H^0_u)\cos\alpha)/\sqrt2$,
and thus couples to the SM particles with
\bea
\frac{g_{ht\bar t}}{g^{\rm SM}_{ht\bar t}} = \frac{\cos\alpha}{\sin\beta}, \quad
\frac{g_{hb\bar b}}{g^{\rm SM}_{hb\bar b}} = -\frac{\sin\alpha}{\cos\beta}, \quad
\frac{g_{h ZZ}}{g^{\rm SM}_{h ZZ}} =
\frac{g_{h WW}}{g^{\rm SM}_{h WW}} = \sin(\beta-\alpha),
\eea
where $g^{\rm SM}_i$ denotes the Higgs coupling in the SM case.
The production of a SM-like Higgs boson at the LHC is dominated by the gluon-gluon
fusion, to which the top quark loop gives the dominant contribution.
Using this property, one can estimate the signal rate for each decay channel of
$H_2=s\sin\theta+h\cos\theta$ compared to the SM prediction:
\bea
R^i_{H_2} =
\frac{\sigma(H_2)\,{\rm Br}(H_2\to i)}{\sigma_{\rm SM}(h)\,{\rm Br}(h\to i)|_{\rm SM}}
\simeq \frac{{\rm Br}(H_2\to i)}{{\rm Br}(h\to i)|_{\rm SM}}
\left(\frac{\cos\alpha}{\sin\beta}\right)^2 \cos^2\theta,
\eea
for small values of $\theta$ and $\delta\equiv(\alpha-\beta+\pi/2)$.
Here the quantities with the subscript, SM, refer to those for the SM case.
For a Higgs boson at 125 GeV, the SM predicts ${\rm Br}(h\to b\bar b) \simeq 0.58$,
${\rm Br}(h\to WW^*) \simeq 0.22$,
${\rm Br}(h\to gg) \simeq 0.09$
and ${\rm Br}(h\to \gamma\gamma) \simeq 2.3\times10^{-3}$
with the total decay width $\Gamma_h\simeq 4.03$ MeV and the production cross section
$\sigma(pp\to h)\simeq 15.3$ pb  \cite{Djouadi:2005gi}.
Hence, for $m_{H_2}=125$ GeV and a moderately large $\tan\beta$, one finds
\bea
R^{b\bar b}_{H_2} &\approx& (1-0.6\,\delta \tan\beta)
\left(1-{\rm Br}(H_2\to \chi^0_1\chi^0_1)\right)\cos^2\theta,
\nonumber \\
R^{WW^*}_{H_2} &\approx& (1+1.4\,\delta \tan\beta)
\left(1-{\rm Br}(H_2\to \chi^0_1\chi^0_1)\right)\cos^2\theta,
\nonumber \\
R^{\gamma\gamma}_{H_2} &\approx& (1+1.4\,\delta \tan\beta)
\left(1-{\rm Br}(H_2\to \chi^0_1\chi^0_1)\right)\cos^2\theta,
\eea
assuming that the Higgs invisible decay is weak and the mixing angles $\theta$
and $\delta$ are small.
Though being a naive estimation, the above shows the general property
of the NMSSM that, if $\delta$ is positive, the mixing between doublet Higgs bosons
enhances the signal rate for $\gamma\gamma$ and $WW^*$ while reducing that for
$b\bar b$ \cite{doublet-mixing}.
In the case that the lightest Higgs boson is singlet-like, $\delta$ is determined
mainly by the mass mixing between doublet Higgs bosons $h$ and $h^\prime$,
\bea
\label{delta}
\delta \simeq \frac{m^2_Z-\lambda^2 v^2}{4|B\mu|}\sin2\beta\sin4\beta
\approx
-\frac{2(m^2_Z-\lambda^2 v^2)}{|B\mu|}
\frac{1}{\tan^2\beta},
\eea
for $m^2_{h^\prime} \simeq 2|B\mu|/\sin2\beta\gg m^2_W$, because the contribution from
the $s^2$-term to $\delta$ requires both $s$-$h$ and $s$-$h^\prime$ mixings and is
further suppressed by the mass ratio, $m^2_S/(2|B\mu|/\sin2\beta)$.
The above indicates that the NMSSM contribution can make $\delta$ positive
when $\lambda^2v^2>m^2_Z$, i.e. when $\lambda$ is larger than about 0.52.
Such effect can be sizable if $h^\prime$ has a mass not far above the weak scale.
Meanwhile, doublet mixing decreases the mass of $H_2$ by a small amount,
\bea
\Delta m_{H_2}|_\delta \simeq
-\frac{(m^2_Z-\lambda^2 v^2)\sin4\beta}{4m_{H_2}}\,\delta,
\eea
for $|\delta|\ll 1$.
However, the tree-level relation $M^2_{hh}|_{\rm tree}=m^2_Z+(\lambda^2v^2-m^2_Z)\sin^22\beta$
implies that $M_{hh}$ is larger than $m_Z$ at the tree-level for $\lambda^2v^2>m^2_Z$.
Thus, when $\delta$ is positive, the decrease of the Higgs mass by the doublet mixing
can be compensated by the NMSSM contribution to the Higgs quartic coupling.

Let us now see the situation in the PQ-NMSSM, where $\lambda\gtrsim0.52$ requires
$\mu\gtrsim 260$ GeV for a moderately large $\tan\beta$ in order to satisfy the
constraint on the invisible $Z$ decay rate.
Provided ${\rm arg}(A_\lambda)={\rm arg}(B_\kappa)$
or ${\rm arg}(A_\lambda)={\rm arg}(B_\kappa)\pm \pi$, all the couplings in the Higgs
potential (\ref{Higgs-V}) can be made real without loss of generality
through an appropriate field redefinition.
In the discussion so far, we have for simplicity assumed
${\rm arg}(A_\lambda)={\rm arg}(B_\kappa)$, for which CP is not spontaneously broken
in the Higgs sector because one can always take a basis where all the couplings
involved are real and positive.
Also note that, in this case, $B$ is larger than $A_\lambda$ as follows from
the minimization condition, and $A_\lambda\sim \mu\tan\beta$ is needed to avoid too
large singlet-doublet mixing.
Hence, when $H_1$ is singlet-like, the heavier doublet Higgs will be very heavy
for $\lambda\gtrsim 0.52$ and $\mu\gtrsim 260$ GeV, making the doublet mixing effect
small.
However, the situation changes for ${\rm arg}(A_\lambda)={\rm arg}(B_\kappa)\pm \pi$.
In this case, depending on the model parameters, the minimum of the Higgs potential
can still lie on a point preserving CP.
Then, since $B$ is not necessarily larger than $A_\lambda$,
it is possible for the heavier doublet Higgs to get a relatively small mass so that
the doublet mixing can yield a sizable positive $\delta$.

\section{Conclusions}

We have examined the implications of singlet-doublet Higgs mixing on the properties
of a SM-like Higgs boson in the PQ-NMSSM, which incorporates the PQ symmetry solving
the strong CP problem and does not suffer from the tadpole and domain-wall problems.
For the case where the lightest Higgs boson is dominated by the singlet scalar,
the Higgs mixing increases the mass of the SM-like Higgs boson while reducing its
couplings to SM particles.
Such mixing effect can be sizable also for large $\tan\beta$ and small $\lambda$,
where the NMSSM direct contribution to the Higgs mass is negligible.
However, the amount of mass increase is limited by the LEP constraints on the properties
of the singlet-like Higgs boson and the lightest neutralino constituted mainly
by the singlino.
In addition, in order for the Higgs signal rate to be close to what the SM predicts,
the mixing should be small.
Combining these, we find that the mixing can enhance the Higgs boson mass by a few GeV or
more even when the ratio for the Higgs signal rate compared to the SM case is to be 0.9.
For the ratio around 0.7, the amount of increase can be as large as about 7 GeV for
the singlet-like Higgs boson around 95 GeV.
Thus, the singlet-doublet Higgs mixing may be crucial for achieving a 125 GeV Higgs mass,
as hinted by the recent ATLAS and CMS data, within the supersymmetric SM with
superparticles having masses around a TeV.
Once the Higgs production cross section is measured, we will learn the bound on how
much the Higgs mixing can modify the properties of the SM-like Higgs boson.

\acknowledgments

We thank Marek Olechowski for pointing out a sign error in the doublet mixing
parameter in an earlier version of the paper, which has been corrected as given
by eq. (\ref{delta}) of the present version.
This work was supported by Grants-in-Aid for Scientific Research from
the Ministry of Education, Science, Sports, and Culture (MEXT),
Japan, No. 23104008 and No. 23540283.

\section*{Low energy effective Higgs potential}

In this appendix, we present the relations between
$(\lambda,m^2_S,A_\lambda,B,\mu,\tan\beta)$ and the SUSY breaking parameters
involved in the Higgs potential.
We also discuss how to integrate out the heavy Higgs scalars in the decoupling
limit.

The SUSY breaking parameters are written
\bea
B_\kappa &=& \frac{1}{B-A_\lambda}\left(
m^2_S + \left(1-\frac{\tan\beta}{\tan^2\beta+1}\frac{A_\lambda}{\mu} \right)
\lambda^2 v^2 \right),
\nonumber \\
m^2_{H_u} &=&
\frac{B\mu}{\tan\beta} - \frac{2 \mu^2 +\lambda^2 v^2}{\tan^2\beta+1}
-\frac{\tan^2\beta-1}{\tan^2\beta+1}\left(\frac{m^2_Z}{2}+\mu^2 \right),
\nonumber \\
m^2_{H_d} &=&
\frac{B\mu}{\cot\beta} - \frac{2 \mu^2 +\lambda^2 v^2}{\cot^2\beta+1}
-\frac{\cot^2\beta-1}{\cot^2\beta+1}\left(\frac{m^2_Z}{2}+\mu^2 \right),
\eea
which are obtained by using the extremum conditions.

Let us construct an effective Higgs potential by integrating out the heavy
Higgs scalars in the decoupling limit
$(2B\mu/\sin2\beta)^{1/2}\gg m_W$.
In the field basis $H=-H_d \sin\alpha+H^c_u \cos\alpha$ and
$H^\prime = H_d \cos\alpha + H^c_u \sin\alpha$ with
$H^c_u=i\sigma_2 H^*_u$, the Higgs potential reads
\bea
V = V_0(S,|H|^2) +  f_2 |H^\prime|^2 - \Big\{
(f_1 \cos^2\alpha- f^*_1\sin^2\alpha)
H^\dagger H^\prime + {\rm h.c.}
\Big\} + \cdots,
\eea
where $f_{1,2}$ are given by
\bea
f_2 &=& m^2_{H_d} \cos^2\alpha + m^2_{H_u} \sin^2\alpha + \lambda^2 |S|^2
- \frac{1}{2} \lambda (A_\lambda S + m^2_0 + {\rm h.c.} ) \sin2\alpha,
\nonumber \\
f_1 &=& \frac{1}{2}(m^2_{H_d}-m^2_{H_u})\tan2\alpha + \lambda (A_\lambda S + m^2_0)
- \frac{1}{4}(g^2+g^{\prime 2}-2\lambda^2)|H|^2\sin2\alpha,
\eea
and the ellipsis indicates Higgs quartic terms with two or more powers
of $H^\prime$.
It is straightforward to find
\bea
\langle f_2 \rangle &=& \frac{2B\mu}{\sin2\beta}\sin^2(\alpha-\beta)
+ {\cal O}(v^2),
\nonumber \\
\langle f_1 \rangle &\propto& \cos(\alpha-\beta).
\eea
Therefore, $\langle f_1 \rangle$ vanishes at $\alpha=\beta-\pi/2$, for which
$\langle H^0 \rangle =v$ and $\langle H^\prime \rangle =0$.
In this case, if $2B\mu/\sin2\beta\gg v^2$, the heavy Higgs doublet $H^\prime$
can be integrated out by solving $\partial_{H^\prime} V =0$:
\bea
\label{sol}
H^\prime = \frac{\sin2\beta}{2B\mu}(f^*_1 \sin^2\beta-f_1\cos^2\beta)
(1+\cdots)H,
\eea
where the ellipsis in the bracket includes terms depending on $S$ and $|H|^2$,
which are irrelevant to the low energy physics because $f_1$ vanishes at the
vacuum.
The relevant interaction terms arise only through the dependence of $f_1$ on
$S$ and $|H|^2$.
Using the extremum conditions, one can rewrite $f_1$,
\bea
f_1 = A_\lambda (\lambda S -\mu)
+ \frac{\sin2\beta}{4}(g^2+g^{\prime 2}-2\lambda^2)(|H|^2-v^2)
\equiv f_{\rm mix}.
\eea
Finally, substituting $H^\prime$ by the solution (\ref{sol}) leads to the effective
potential (\ref{Higgs-potential}).

\end{document}